\documentclass[a4paper,aps,prl,floatfix,twocolumn,footinbib,showpacs,superscriptaddress]{revtex4-1}

\usepackage[pdftex]{graphicx}
\setkeys{Gin}{width=0.8\columnwidth}
\usepackage{latexsym,amsmath}
\usepackage[pdftex]{color}
\usepackage{hyperref}
\usepackage{float}
\usepackage[all]{xy}
\usepackage{verbatim}

\newcommand{\av}[1]{\langle #1 \rangle}

\begin{document}

\title{Effects of heterogeneous social interactions on flocking
  dynamics}

\author{M. Carmen Miguel}

\affiliation{Departament de F\'{\i}sica de la Mat\`{e}ria Condensada,
  Universitat de Barcelona, Mart\'{\i} i Franqu\`es 1, 08028 Barcelona,
  Spain}

\affiliation{Universitat de Barcelona Institute of Complex Systems
  (UBICS), Universitat de Barcelona, Barcelona, Spain}

\author{Jack T. Parley}

\affiliation{Departament de F\'isica, Universitat Polit\`ecnica de
  Catalunya, Campus Nord B4, 08034 Barcelona, Spain}

\author{Romualdo Pastor-Satorras}

\affiliation{Departament de F\'isica, Universitat Polit\`ecnica de
  Catalunya, Campus Nord B4, 08034 Barcelona, Spain}

\date{\today}

\begin{abstract}
  Social relationships characterize the interactions that occur within
  social species and may have an important impact on collective animal
  motion.  Here, we consider a variation of the standard Vicsek model
  for collective motion in which interactions are mediated by an
  empirically motivated scale-free topology that represents a
  heterogeneous pattern of social contacts. We observe that the degree
  of order of the model is strongly affected by network heterogeneity:
  more heterogeneous networks show a more resilient ordered state; while
  less heterogeneity leads to a more fragile ordered state that can be
  destroyed by sufficient external noise. Our results challenge the
  previously accepted equivalence between the {\em static} Vicsek model
  and the equilibrium XY model on the network of connections, and point
  towards a possible equivalence with models exhibiting a different
  symmetry.
\end{abstract}

\maketitle

Collective motion in living and complex systems~\cite{Vicsek2012}, where
simple interactions between constituent entities produce striking
spatio--temporal patterns on scales larger than the entities themselves,
are commonplace.  Some of the examples that best highlight the emergence
of such patterns are found in animal
motion~\cite{Sumpter2006,Sumpter10}, where the animals collectively
exhibit some of the most spectacular and fascinating sights in
nature. These include flocks of birds turning in unison or migrating in
well-ordered formation, shoals of fish splitting and reforming as they
outmaneuver a predator, seasonal migratory herds of large herbivores,
etc.

The challenge of understanding how hundreds or thousands of organisms
move together and give rise to such intriguing collective responses in
the absence of any apparent leader or driving field has attracted the
attention of the scientific community for a long time. Significant
progress in understanding how some of these features come about has been
achieved through the development of relatively simple models of
self-propelled particles (SPP). In SPP models, the complex dynamics of
individuals within a group are simplified to those of particles that
move with given velocities and experience flocking interactions within a
local interaction zone, combined with random fluctuations due to
intrinsic or environmental factors. In the celebrated Vicsek
model~\cite{Vicsek1995}, these interactions consist in
the alignment of the velocity of an SPP with the average velocity of
some of its neighbors.  Perfect alignment is, however, impeded by the
addition of a noise term that mimics, for instance, the difficulties in
gathering and processing the surrounding information. The success of the
model lies in the production of a phase transition as a function of
noise intensity, $\eta$, separating an ordered or polarized (flocking)
phase at $\eta \leq \eta_c$, where particles travel in a common
direction, from a disordered phase for $\eta > \eta_c$, where particles
behave as uncorrelated persistent random
walkers~\cite{Ginelli2016,mendez14}. This is particularly fruitful due
to the analogies that can be drawn between the self-organization of
herds of moving animals and standard phase transitions observed in
condensed matter~\cite{Vicsek2012,yeomans}.

The main assumption of the Vicsek and other similar models of collective
motion \cite{Couzin02,Chate2008} is that particles tend to orient their
velocity parallel to the average velocity in a local neighborhood,
independently of their identity.  This kind of interaction rule leaves
aside, however, the important fact that real interactions between moving
animals can be more intricate. One source of complication can be the
presence of \textit{social interactions}~\cite{croft2008exploring}
between the group members, which can lead, in the framework of the
Vicsek model, to a tendency to align one's velocity with that of
individuals with which one has strong social ties, but that might be
separated by a relatively long Euclidean distance. The presence of such
social interactions, naturally represented in terms of social
networks~\cite{Newman10}, has been observed in
mammals~\cite{LusseauS477,Flack:2005aa,Rhodes06} and
fish~\cite{CroftS516,Croft2005}, and has even been studied in the
context of schooling fish~\cite{Rosenthal2015}.

The impact of social interactions given in terms of networks has already
been considered in the context of collective motion and the Vicsek
model~\cite{Aldana2003,Aldana2007,Pimentel2008,SEKUNDA2016,Bode2011,Bode2011a};
but, to the best of our knowledge, an in-depth study is still lacking.
Here, we focus on the effects of the topological heterogeneity observed
in certain animal social networks~\cite{Lusseau2003,Manno2008}, which
can be represented by a degree distribution $P(k)$, defined as the
probability that a randomly chosen individual is connected to $k$ other
individuals, showing a scale-free signature~\cite{Barabasi:1999} of the
form: $P(k) \sim k^{-\gamma_d}$. We study the behavior of the Vicsek
model when applied to complex networks with varying heterogeneity (a
varying degree exponent $\gamma_d$), generated using the uncorrelated
configuration model (UCM)~\cite{Catanzaro05}. In this setting, each
particle's neighbors always remain the same.  As a consequence, it is
usually assumed that, in this limit, the Vicsek model must be equivalent
to the equilibrium XY model of ferromagnetism defined on the network of
connections (see e.g.  Refs.~\cite{Vicsek1995,Czirok1997,Ginelli2016}).
The XY model has been theoretically and numerically analyzed in
scale-free networks under various
conditions~\cite{dorogovtsev07:_critic_phenom,Kwak07,Yang2008}.  By
means of extensive numerical simulations, we show that on static
scale-free networks the Vicsek and the XY models exhibit
\textit{different} critical behavior.  Furthermore, our simulations are
compatible with the behavior reported for the non-equilibrium
majority-vote model with noise applied to complex
networks~\cite{Chen2015}.

We consider a version of the Vicsek model in which interactions are
mediated by a static complex network, with links representing social
interactions. A social network can be fully represented in terms of its
adjacency matrix $a_{ij}$~\cite{Newman10}, with value $a_{ij} = 1$, if
individuals $i$ and $j$ are socially connected, while $a_{ij} = 0$
otherwise.  By considering an ordering dynamics based on social
interactions alone, we disregard spatial position, and thus the SPPs are
uniquely specified in terms of their velocity $\mathbf{v}_i(t)$, assumed
to be normalized: $|\mathbf{v}_i(t)| = v_0$.  We fix $v_0=1$. We
consider velocities in two dimension, fully determined by the angle
$\theta_i(t)$ they form with, say, the $x$ axis, i.e.
$\mathbf{v}_i(t) = \{ \cos \theta_i(t), \sin \theta_i(t) \}$.  With the
original definition of the model~\cite{Ginelli2016}, velocities are
synchronously updated via the rule:
\begin{equation}
  \label{eq:3}
  \theta_i(t+1) = \Theta\left[ \mathbf{v}_i(t) + \sum_{j=1}^N a_{ij}
    \mathbf{v}_j(t) \right] + \eta \; \xi_i(t), 
\end{equation}
where $N$ is the network size, $\Theta[\mathbf{V}]$ represents the angle
described by vector $\mathbf{V}$, $\xi_i(t)$ is random noise
uniformly distributed within the interval $[-\pi, \pi]$, and
$\eta \in [0,1]$ is a parameter that reflects the noise strength. Note that
$\eta = 1$ is the maximum possible noise, since it corresponds to a
completely disordered system. 

The phase transition between ordered and disordered states in the Vicsek
model is determined by the temporal evolution of an order parameter
$\phi_\eta(t)$, defined as~\cite{Vicsek1995}:
\begin{equation}
  \label{eq:5}
  \phi_\eta(t) = \frac{1}{N} \left| \sum_{i=1}^N \mathbf{v}_i(t) \right|.
\end{equation}
From here, one defines the average order parameter
$\av{\phi_\eta} = \lim_{T\to\infty} \frac{1}{T} \int_0^T \phi_\eta(t) \;
dt$ and the susceptibility
$\chi_\eta = N [\av{\phi_\eta^2} - \av{\phi_\eta}^2]$, which close to
the critical point behave as
$\av{\phi_\eta} \sim {(\eta_c - \eta)}^\beta$ and
$\chi_\eta \sim {|\eta_c - \eta|}^{-\gamma}$, respectively, defining the critical
exponents $\beta$ and $\gamma$, in analogy with the ferromagnetic phase
transition~\cite{yeomans}. 

The model defined by Eq.~(\ref{eq:3}) does not admit a feasible
analytical treatment for general networks~\footnote{See however
  Ref.~\cite{Pimentel2008} for a solution in the case of random regular
  (homogeneous) networks.}. We can, however, solve it in the fully
connected case. To proceed, it is convenient to write the order
parameter in the alternative form:
\begin{equation}
  \label{eq:1}
  \phi_\eta(t) = \frac{1}{N} \sum_{i=1}^N \cos \left[ \theta_i(t) -
    \bar{\theta}(t)\right],
\end{equation}
where $\bar{\theta}(t) = \Theta[\sum_{i=1}^N
\mathbf{v}_i(t)]$. Eq.~(\ref{eq:1}) can be shown to be exactly equal to
Eq.~(\ref{eq:5}), see Appendix~\ref{sec:altern-form-order}. For a fully
connected network, the Vicsek model can be solved starting from
Eq.~(\ref{eq:1}) (see Appendix~\ref{sec:analyt-solut-fully}), obtaining
the result that the system is ordered for any $\eta < 1$. In the
vicinity of this point, expansions of the solution lead to
$ \av{\phi_\eta} \sim 1- \eta$ and $\chi_\eta \sim \mathrm{const.}$,
leading to the critical exponents $\beta = 1$, $\gamma = 0$.

In the case of sparse networks, it is usually assumed that, when the
particles are immobile and the network of connections is sufficiently
dense, the Vicsek model can be mapped to the equilibrium XY
model~\cite{Ginelli2016}, where the temperature $T$ is a function of the
noise intensity $\eta$, fulfilling the limits $T \to 0$ for
$\eta \to 0$, and $T \to \infty$ for $\eta \to 1$. The XY model applied
to networks can be solved within an annealed network approximation,
obtaining a critical temperature
$T_c = J \av{k^2} /[2 \av{k}]$~\cite{dorogovtsev07:_critic_phenom},
where $J$ is the coupling constant of the XY Hamiltonian. That is, for
scale-free networks with $\gamma_d>3$, there is a true transition at a
finite critical temperature; while for $\gamma_d\leq 3$, there is no
transition and the system is always ordered for any finite $T$. These
results have been confirmed by numerical simulations on
heterogeneous~\cite{Yang2008} and homogeneous~\cite{PhysRevE.67.036118}
networks.

\begin{figure}[t]
  \includegraphics{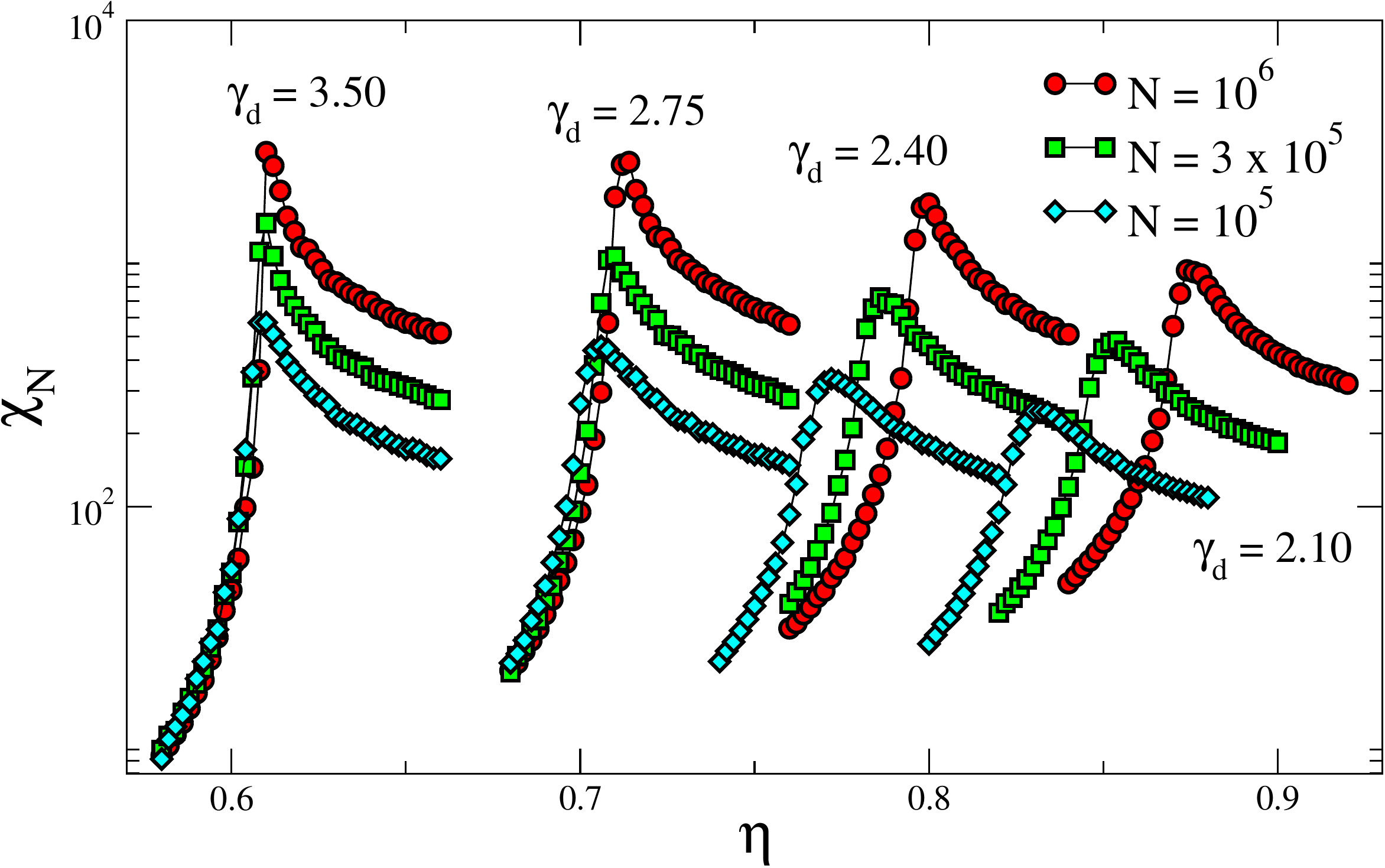}
  \caption{Numerical dynamic susceptibility as a function of noise
    amplitude, $\eta$, in UCM networks of different size. The groups of
    functions for different $N$ correspond to the values of $\gamma_d$
    (from left to right): $3.50$, $2.75$, $2.50$ and $2.10$.}
  \label{fig:eta_FSS}
\end{figure}

\begin{figure}[t]
 \includegraphics{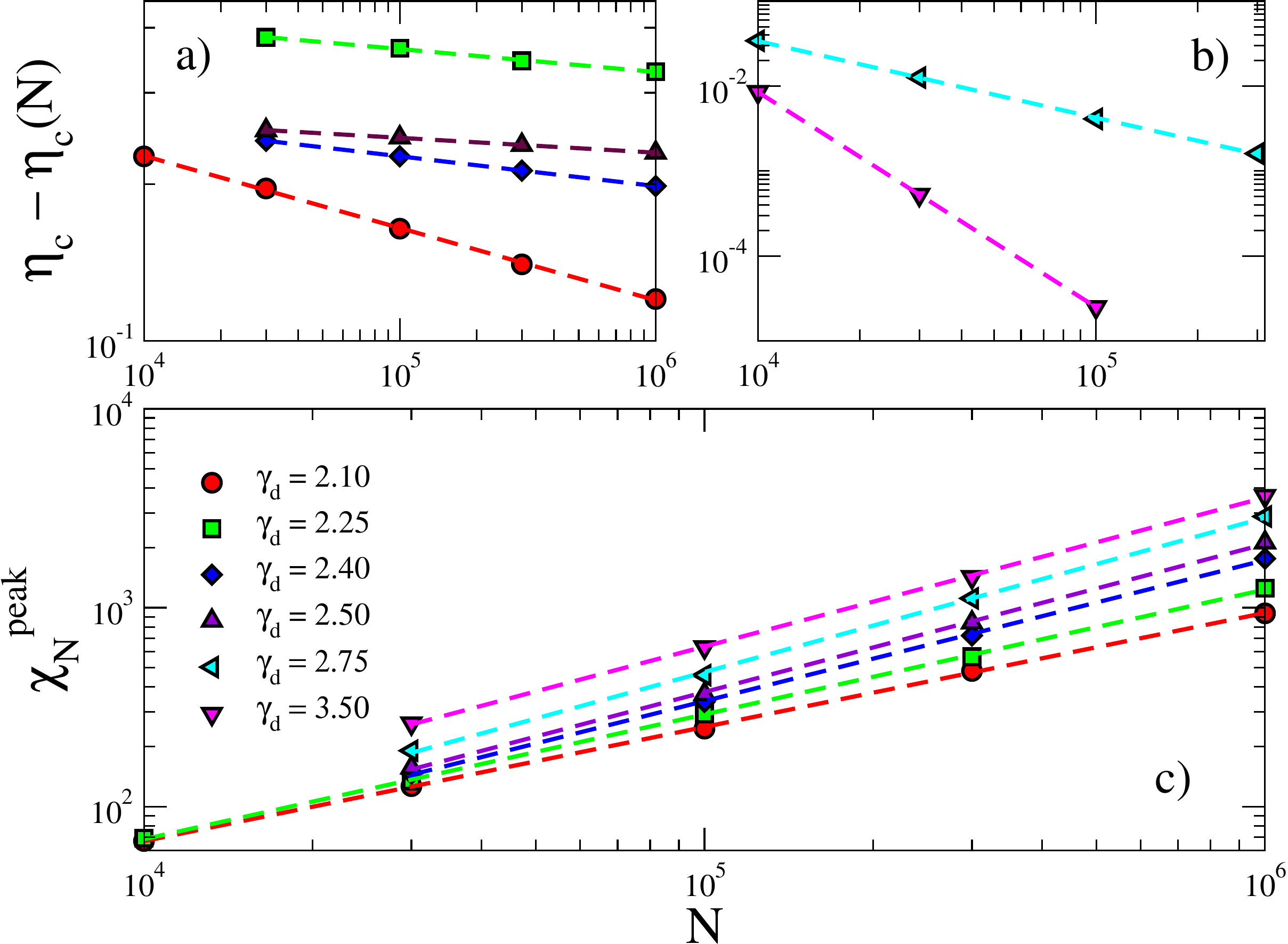}
 \caption{Finite size scaling analysis of the position of the
   susceptibility peak, according to Eq.~(\ref{eq:14}) for small (a) and
   large (b) values of the degree exponent $\gamma_d$. The critical
   points, in the thermodynamic limit, are given
   Table~\ref{tab:values}. c) Scaling of the peak of susceptibility with
   network size for different values of $\gamma_d$. The associated
   exponents $\chi_n^\mathrm{peak} \sim N^\delta$, with
   $\delta = (\beta + \gamma)/\nu$, are given in
   Table~\ref{tab:values}.}
  \label{fig:eta_thermodynamic}
\end{figure}

In order to check the validity of the mapping to the XY model, we
performed numerical simulations of the Vicsek model on UCM networks with
different values of $\gamma_d$ and a minimum degree of $m=3$
\cite{Catanzaro05}. The order parameter, $\av{\phi_\eta}$, is computed
by averaging over $50,000$ time steps, after letting the system
initially relax for $10,000$ time steps.
\begin{figure}[t]
  \includegraphics{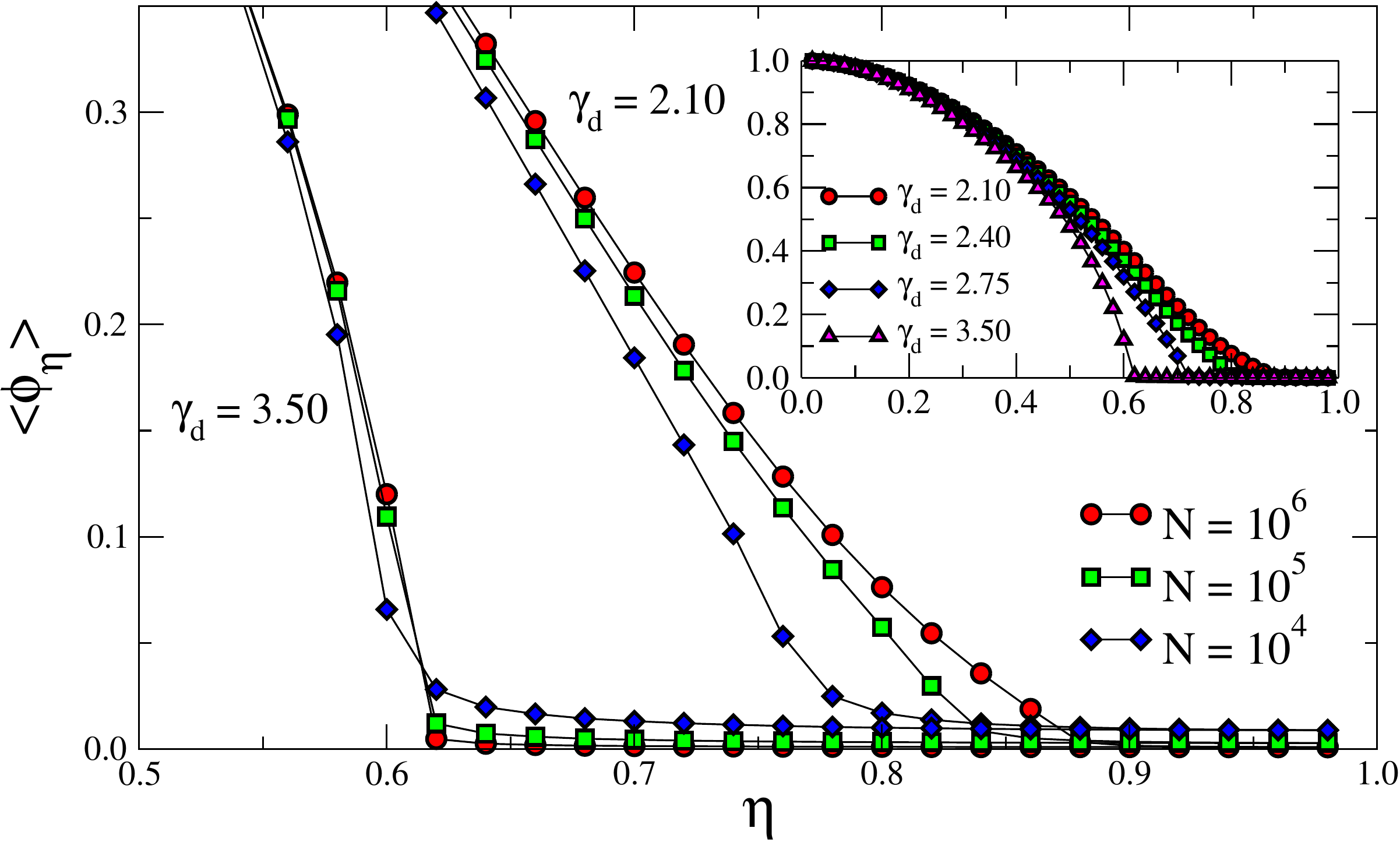}
  \caption{Inset: Average order parameter as a function of the noise
    intensity, $\eta$, for different values of the degree exponent
    $\gamma_d$ in UCM networks of size $N=10^6$. Main: The order parameter as
    a function of $\eta$ for different values of the network size,
    $N$. The sets of plots correspond to $\gamma_d = 2.1$ (left) and
    $\gamma_d = 3.5$ (right).}
  \label{fig:rho_vs_eta_2}
\end{figure}
Fig.~\ref{fig:rho_vs_eta_2}(inset) shows a plot of the average order
parameter as a function of $\eta$, computed in networks of size
$N = 10^6$ with different degree exponents.  Fig.~\ref{fig:rho_vs_eta_2}
(main) illustrates the effects of system size for two different values
of $\gamma_d$.  From this figure it is apparent that for small
$\gamma_d$ the effective threshold depends strongly on $N$.  In order to
explore size effects in greater detail, we proceed to compute the
effective threshold by looking at the \textit{dynamic susceptibility}:
\begin{equation}
  \label{eq:13}
  \chi_N(\eta) = N \frac{\av{\phi_\eta^2} -
    \av{\phi_\eta}^2}{\av{\phi_\eta}}, 
\end{equation}
which is customarily used to detect phase transitions in complex
networks~\cite{Ferreira2012,Castellano2016}. The effective critical
point, $\eta_c(N)$, will be given, for a given network size $N$, by the
position of the \textit{maximum} of the dynamic susceptibility
$\chi_N(\eta)$.  The critical point in the thermodynamic limit
$N\to\infty$ can be obtained by applying a finite-size scaling
hypothesis~\cite{cardy88} of the form:
\begin{equation}
  \label{eq:14}
  \eta_c(N) = \eta_c - a N^{-1/\nu},
\end{equation}
where $\nu$ is another characteristic critical exponent
\cite{Binder2010,Ferreira2012}. The height of the peak of the dynamic
susceptibility, $\chi_N^\mathrm{peak}$, also scales with $N$, adopting the 
form~\cite{Ferreira2012,Castellano2016}:
\begin{equation}
  \label{eq:6}
  \chi_N^\mathrm{peak} \sim N^{(\beta + \gamma)/\nu}.
\end{equation}
In Fig.~\ref{fig:eta_FSS} we plot the dynamic susceptibility,
$\chi_N(\eta)$, for networks with different values of the degree
exponent $\gamma_d$.
As can be seen from the figure, for $\gamma_d > 2.5$, the location of
the peak of the susceptibility appears to tend to a constant value
smaller than $1$. In contrast, for $\gamma_d < 2.5$, this location
shifts to larger values of $\eta$ as $N$ increases. We proceed to
estimate the critical point in the thermodynamic limit by applying a
non-linear fit to the position of the peak, $\eta_c(N)$, as a function
of $N$, according to Eq.~(\ref{eq:14}); see
Fig.~\ref{fig:eta_thermodynamic} and Table~I.

\begin{table}[b]
  \begin{ruledtabular}
    \begin{tabular}{|l|cccccc|}
      $\gamma_d$ & $2.10$ & $2.25$ & $2.40$ & $2.50$ & $2.75$ & $3.50$
      \\ \hline
      $\eta_c$  & $0.99(1)$ & $1.00(5)$ & $1.00(5)$ & $1.00(1)$ &
                                                                  $0.71(1)$
                                                              &
                                                                $0.61(1)$\\
      $\delta$ & $0.574(3)$ & $0.63(1)$ & $0.71(1)$ & $0.74(2)$ &
                                                                  $0.77(3)$
                                                              & $0.75(1)$
    \end{tabular}
    
  \end{ruledtabular}
\caption{Critical point and exponent $\delta$ for the Vicsek model in
  scale-free networks with different degree exponent.}
\label{tab:values}
\end{table}

From these results, it is apparent that, for $\gamma_d > 2.50$, the
critical point, $\eta_c$, tends to a constant value of less than $1$;
while for $\gamma_d \leq 2.50$, the critical point tends to $1$ in the
thermodynamic limit. Therefore, in this latter case, the order--disorder
transition characteristic of the model is suppressed, and the system is
fully ordered for any $\eta<1$ in sufficiently large networks. In
contrast, for $\gamma_d > 2.50$, there is a true order--disorder
transition, which is preserved even in the limit of infinite network
size. While the exponent $\nu$ is difficult to estimate due to
statistical fluctuations in the non-linear fitting procedure, the
exponent $\delta \equiv (\beta + \gamma)/\nu$, controlling the growth of
the dynamic susceptibility peak, can be reliably computed; see
Fig.~\ref{fig:eta_thermodynamic} and Table~I. The exponents obtained are
again compatible with a radical difference in behavior between
$\gamma_d \geq 2.50$, for which we obtain $\delta \simeq 0.75$; and
$\gamma_d<2.50$, where $\delta$ is an increasing function of $\gamma_d$.

The numerical results obtained for heterogeneous scale-free networks
provide a clear picture: when dealing with networks, the Vicsek model
cannot be directly mapped to the XY model. The main evidence of this
incompatibility comes from the behavior of the critical point. For the
XY model, one expects a finite critical temperature (i.e., $\eta_c < 1$)
for $\gamma_d >3$; and an infinite critical temperature (i.e.,
$\eta_c = 1$), or in other words, no phase transition, for
$\gamma_d < 3$. Meanwhile, when applied to networks, the Vicsek model
only produces a true order--disorder transition for degree exponents
larger than $\gamma_d = 5/2$. Experimental characterizations of the
degree exponent in groups of social animals (despite the conceivable
difficulties associated with measuring it and the fact that it probably
varies depending on the behavioral test chosen) provide $\gamma_d$
values within the range $1-3.5$~\cite{Lusseau2003,Manno2008}. It is also
obvious that the thermodynamic limit cannot be achieved in
experiments. Nevertheless, our results could have important consequences
for the resilience of the ordered phase observed in different species,
according to the heterogeneity of their social contact
distribution. Strongly heterogeneous networks show a resilient ordered
phase for the whole range of disorder values; while low heterogeneity
leads to a more fragile ordered phase that can be destroyed by a
sufficient amount of external noise.

In order to shed some light on the behavior observed, we put forward the
following hypothesis: given that in networks, the dimensionality of the
order parameter appears to be irrelevant (for example, the Ising and XY
models share the same scaling of the critical point and the same
critical exponents~\cite{dorogovtsev07:_critic_phenom}), we conjecture
that a model analogous to the Vicsek model, but with a scalar order
parameter, might also share the same behavior as the Vicsek model in
heterogeneous networks.  In this way, we consider the majority-vote
model \cite{Aldana2007}, in which spin variables on the vertices of a
network update their state taking the value of the majority of their
nearest neighbors. This state is randomly flipped with a probability
$f$, which plays a similar role to the noise strength, $\eta$, but takes
a maximum value $f_\mathrm{max} = 1/2$~\cite{liggett97}. On a fully
connected graph, the majority-vote model shows a critical point
$f_c = 1/2$, which in the Vicsek case translates to $\eta_c = 1$ with
exponents $\beta = 1$ and $\gamma=0$, see
Appendix~\ref{sec:majority-vote-model}.  Meanwhile, in heterogeneous
networks with a power-law degree distribution, a threshold
$f_c = 1/2 - \sqrt{\frac{\pi}{8}}\frac{\av{k}}{\av{k^{3/2}}}$ has
recently been reported~\cite{Chen2015}. This threshold shows a
transition from $f_c < 1/2$ for $\gamma_d > 5/2$ to $f_c = 1/2$ for
$\gamma_d < 5/2$ in the thermodynamic limit: in full agreement with the
observations of the Vicsek model applied to networks. Moreover, above
the threshold degree exponent, $\gamma_d = 5/2$, the value of the
exponent $(\beta + \gamma)/\nu \simeq 0.75$ is also in agreement with
the mean-field values of the majority-vote mode: $\beta = 1/2$,
$\gamma = 1$, $\nu = 2$ \cite{PhysRevE.71.016123}. In order to confirm
the equivalence of the majority-vote and Vicsek models on heterogeneous
networks, we have performed additional extensive simulations of the
latter for a range of different degree exponents on UCM networks. The
results obtained are described in
Appendix~\ref{sec:majority-vote-model-1}. From our simulations, we
confirm the results in Ref.~\cite{Chen2015} regarding a threshold
$f_c \to 1/2$ in the thermodynamic limit for $\gamma_d < 5/2$, while
$f_c < 1/2$ for $\gamma_d>5/2$. The estimation of the exponent $\delta$
for the growth of the dynamical susceptibility peak with network size,
Eq.~(\ref{eq:6}), leads to the results $\delta = 0.57(1)$ for
$\gamma_d=2.10$, $\delta=0.61(1)$ for $\gamma_d = 2.25$,
$\delta = 0.67(2)$ for $\gamma_d=2.40$, and $\delta=0.78(2)$ for
$\gamma_d = 2.75$. The excellent agreement of these exponents, compared
with the ones for the Vicsek model reported in Table~\ref{tab:values},
confirm our hypothesis regarding the equivalence of Vicsek and
majority-vote model on complex networks.

We finally  focus on the hierarchy of the order of the nodes of
different degree in the Vicsek model, and compute a degree-restricted
order parameter defined
as
\begin{equation}
  \label{eq:15}
  \phi_\eta(t; k) = \frac{1}{N_k} \sum_{i \in \mathcal{V}_k} \cos
  \left[\theta_i(t) - \bar{\theta}(t) \right],
\end{equation}
where $\mathcal{V}_k$ is the set of nodes with degree $k$, and $N_k$ is
the number of such nodes. From this expression, a time-independent order 
parameter, $\av{\phi_\eta(k)}$, is defined by means of an appropriate time
average over a large time window, $T$. 
\begin{figure}[t]
  \includegraphics{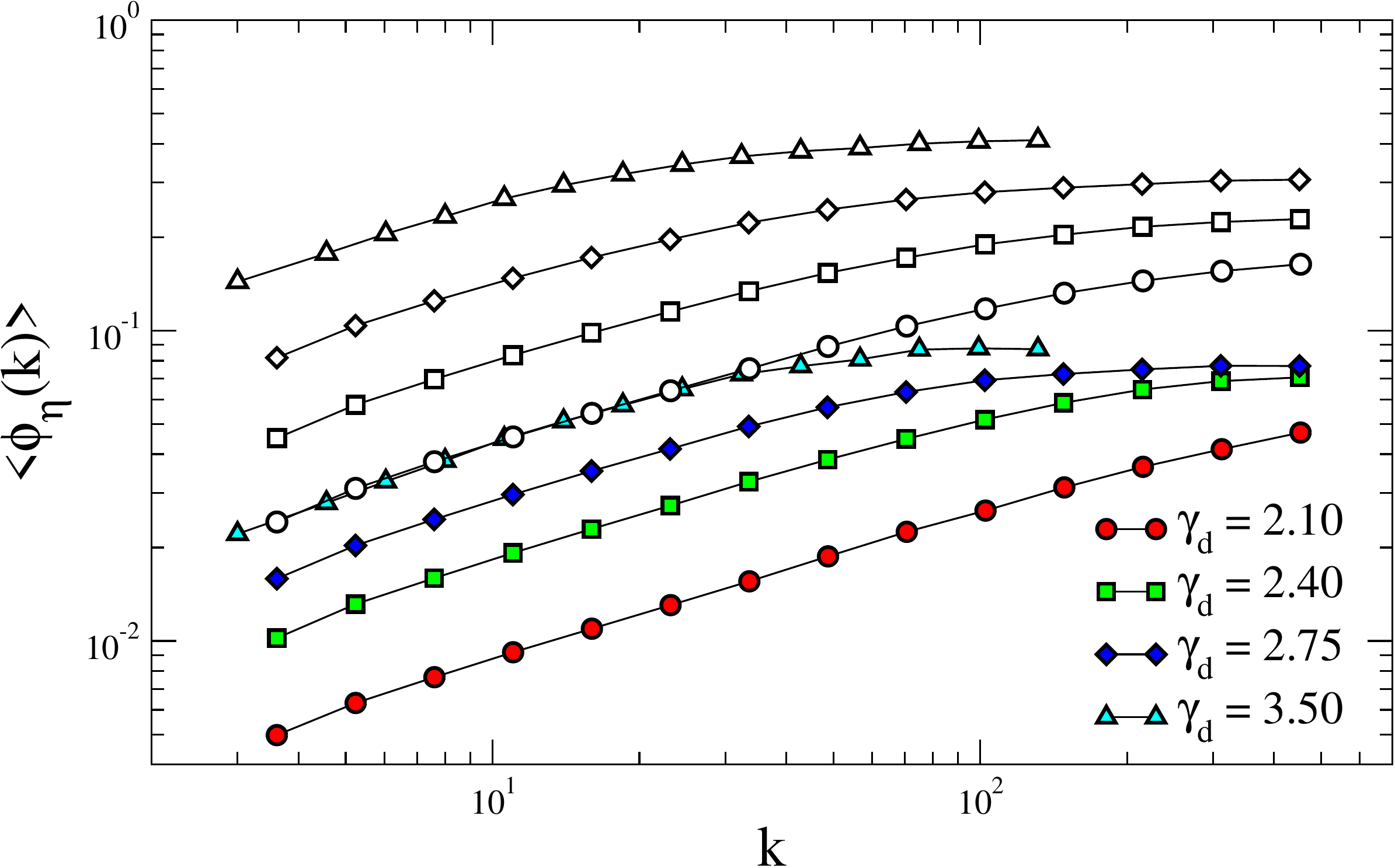}
  \caption{Restricted order parameter as a function of degree for UCM
    networks of size $N=3\times10^5$ and different degree exponent
    values, $\gamma_d$. Solid symbols correspond to a noise strength
    equal to the peak of the dynamic susceptibility; outline symbols
    correspond to a noise strength $3\%$ above the peak.}
  \label{fig:order_k_func}
\end{figure}
In Fig.~\ref{fig:order_k_func}, we plot the restricted order parameter
as a function of $k$. As can be seen, there is apparently a hierarchy in
the order of the systems, with low-degree nodes being more disordered
than high-degree nodes. This can be explained by the larger number of
connections of high degree nodes, which average velocities over a larger
ensemble that low-degree nodes do and are therefore less susceptible to
the influence of the external noise. This effect can be
interpreted as high degree nodes playing the role of leaders, that can
keep the network ordered even close to the maximum possible value of
disorder when they are large enough (i.e., for small values of
$\gamma_d$).

In conclusion, we have studied numerically the Vicsek model applied to
complex scale-free networks with a degree distribution
$P(k) \sim k^{-\gamma_d}$. By means of extensive numerical simulations,
we observe that the nature of the possible order--disorder transition
exhibited by the model depends on the level of heterogeneity of the
network, as given by the value of the degree exponent $\gamma_d$. For
$\gamma_d > 2.5$, there is a true transition, located at
$\eta_c(\gamma_d)$ which increases with decreasing
$\gamma_d$. Meanwhile, for $\gamma_d<2/5$, we obtain a critical point in
the thermodynamic limit equal to $1$, indicating the lack of a true
critical transition.  These results indicate that flocking dynamics in
scale-free social networks is more robust against noise effects in the
case of high network heterogeneity (i.e.~small $\gamma_d$).  These
numerical results are in disagreement with the validity of direct
mapping of the Vicsek model to the equilibrium XY model on the network
of connections, which is usually assumed to be valid.  Nonetheless, our
results do appear to be in agreement with those of the non-equilibrium
majority-vote model on complex networks, which can be considered as a
variation of the Vicsek model with reduced symmetry of the order
parameter.  Our work highlights the role of the effects of social
topology in flocking dynamics and open up intriguing questions regarding
the role of symmetries in dynamical processes on networks. Deeper
research effort in necessary to further our understanding of both
questions.

\begin{acknowledgments}
  We acknowledge financial support from the Spanish Government's MINECO, under
  projects FIS2013-47282-C2-1-P, FIS2013-47282-C2-2,
  FIS2016-76830-C2-1-P and FIS2016-76830-C2-2-P. Also, R. P.-S. acknowledges
  additional financial support from ICREA Academia, funded by the
  \textit{Generalitat de Catalunya} regional authorities.
\end{acknowledgments}

\appendix

\section{Alternative form of the order parameter}
\label{sec:altern-form-order}

In order to show that the proposed alternative form of the order
parameter for the Vicsek model is identical to the original one, we
start from the expression
\begin{equation}
  \label{eq:1}
  \phi_\eta(t) = \frac{1}{N} \sum_{i=1}^N \cos \left[ \theta_i(t) -
    \bar{\theta}(t)\right].
\end{equation}
Using the cosine angle difference identity, we can write
\begin{equation}
  \label{eq:8}
  \phi_\eta(t) = \frac{1}{N} \sum_{i=1}^N \left[ \cos  \theta_i(t) \cos
    \bar{\theta}(t)  + \sin  \theta_i(t) \sin
    \bar{\theta}(t)\right].
\end{equation}
Now, using
\begin{eqnarray}
  \label{eq:9}
  \cos \bar{\theta}(t) &=& \frac{\sum_i \cos \theta_i(t)}{\sqrt{
  \left(\sum_i \cos \theta_i(t)\right)^2 + \left(\sum_i \sin
  \theta_i(t)\right)^2}}, \\
 \sin \bar{\theta}(t) &=& \frac{\sum_i \sin \theta_i(t)}{\sqrt{
  \left(\sum_i \cos \theta_i(t)\right)^2 + \left(\sum_i \sin
                          \theta_i(t)\right)^2}},
\end{eqnarray}
and substituting into Eq.~(\ref{eq:8}), we obtain
\begin{equation}
  \label{eq:10}
   \phi_\eta(t) = \frac{1}{N} \sqrt{
  \left(\sum_i \cos \theta_i(t)\right)^2 + \left(\sum_i \sin
                          \theta_i(t)\right)^2},
\end{equation}
which is exactly the form of the temporal order parameter in its
classical definition, Eq.~(2) in the main paper.

\section{Analytical solution for fully connected networks}
\label{sec:analyt-solut-fully}

In the case of fully connected networks, in which every node is
connected to every other one, we have
$\theta_i(t) = \bar{\theta}(t) + \eta \xi_i(t)$, an therefore
\begin{equation}
  \label{eq:11}
  \phi_\eta(t) = \frac{1}{N} \sum_{i=1}^N \cos \left[ \eta
    \xi_i(t)\right]. 
\end{equation}
As $\xi_i(t)$ is uncorrelated in both the particle index an time, we can
write the average value, in the thermodynamic limit,
\begin{equation}
  \label{eq:12}
  \av{\phi_\eta}  = \frac{1}{2 \pi} \int_{-\pi}^{\pi} \cos \left[ \eta
    \xi\right] \; d \xi = \frac{\sin(\eta \pi)}{\eta \pi}.
\end{equation}
This value of the average order parameter is different from zero for any
$\eta < 1$, indicating $\eta_c = 1$. In the vicinity of this critical
point, a Taylor expansion of Eq.~(\ref{eq:12}) leads to $\av{\phi_\eta}
\sim 1-\eta$, defining the critical exponent $\beta = 1$.

To compute the susceptibility, $\chi_\eta = N [\av{\phi_\eta^2} -
\av{\phi_\eta}^2]$, we start from the variance of $\phi_\eta$
\begin{eqnarray}
  \label{eq:1}
  \mathrm{Var}(\phi_\eta) &=&
   \left< \frac{1}{N^2} \sum_{i=1}^N
   \cos(\eta \xi_i) \sum_{j=1}^N  \cos(\eta \xi_j) \right> \nonumber 
                              -\left(  \frac{\sin(\eta \pi)}{\eta \pi}\right)^2 \\
  &=& \left< \frac{1}{N^2}  \sum_{i=1}^N
   \cos^2(\eta \xi_i)  + \frac{1}{N^2} \sum_{i \neq j}^N \cos(\eta
      \xi_i) \cos(\eta \xi_j)\right> \nonumber \\
                          &-& \left(  \frac{\sin(\eta \pi)}{\eta \pi}\right)^2 \nonumber\\
                          &=& \frac{1}{2\pi N} \int_{-\pi}^\pi\cos^2(\eta \xi) d\xi
                              + \frac{N-1}{2\pi N} \left[
                              \int_{-\pi}^\pi\cos(\eta \xi)
                              d\xi\right]^2 \nonumber \\
                          &-& \left(  \frac{\sin(\eta \pi)}{\eta \pi}\right)^2 \nonumber\\
  &=& \frac{1}{4N} \left(2+ \frac{\sin(2\eta \pi)}{\eta \pi} \right) -
      \frac{1}{N} \left(  \frac{\sin(\eta \pi)}{\eta \pi}\right)^2.
\end{eqnarray}
From here, we obtain the susceptibility
\begin{equation}
  \label{eq:2}
  \chi_\eta  = \frac{1}{2} \left(1+ \frac{\sin(2\eta \pi)}{2 \eta \pi} \right) -
  \left(  \frac{\sin(\eta \pi)}{\eta \pi}\right)^2.
\end{equation}

An expansion of Eq.~(\ref{eq:2}) around the critical point $\eta_c = 1$
leads to $\chi_\eta \sim \frac{1}{2} - (1-\eta)$, yielding the critical
exponent $\gamma = 0$. Therefore, for the dynamic susceptibility
$\chi_\infty(\eta)$, Eq.~(5) in the main paper, we have in the
thermodynamic limit
\begin{equation}
  \label{eq:3}
  \chi_\infty(\eta) \sim \frac{1}{2} (1-\eta)^{-1},
\end{equation}
compatible with $\beta + \gamma = 1$.

\section{Majority vote model on fully connected networks}
\label{sec:majority-vote-model}

In order to explore the behavior of the majority vote model on fully
connected networks we have performed numerical simulations of the
dynamics in which each node takes the majority state of the other $N-1$
nodes with probability $1-f$, and the opposite state with probability
$f$. The order parameter $\av{\phi_f}$ is defined as the average
absolute value of the magnetization in the steady state. The dynamic
susceptibility is, in its turn, defined as
\begin{equation}
  \label{eq:4}
  \chi_N(f) = N \frac{\av{\phi_f^2} - \av{\phi_f}^2}{\av{\phi_f}}. 
\end{equation}

\begin{figure}[t]
  \includegraphics{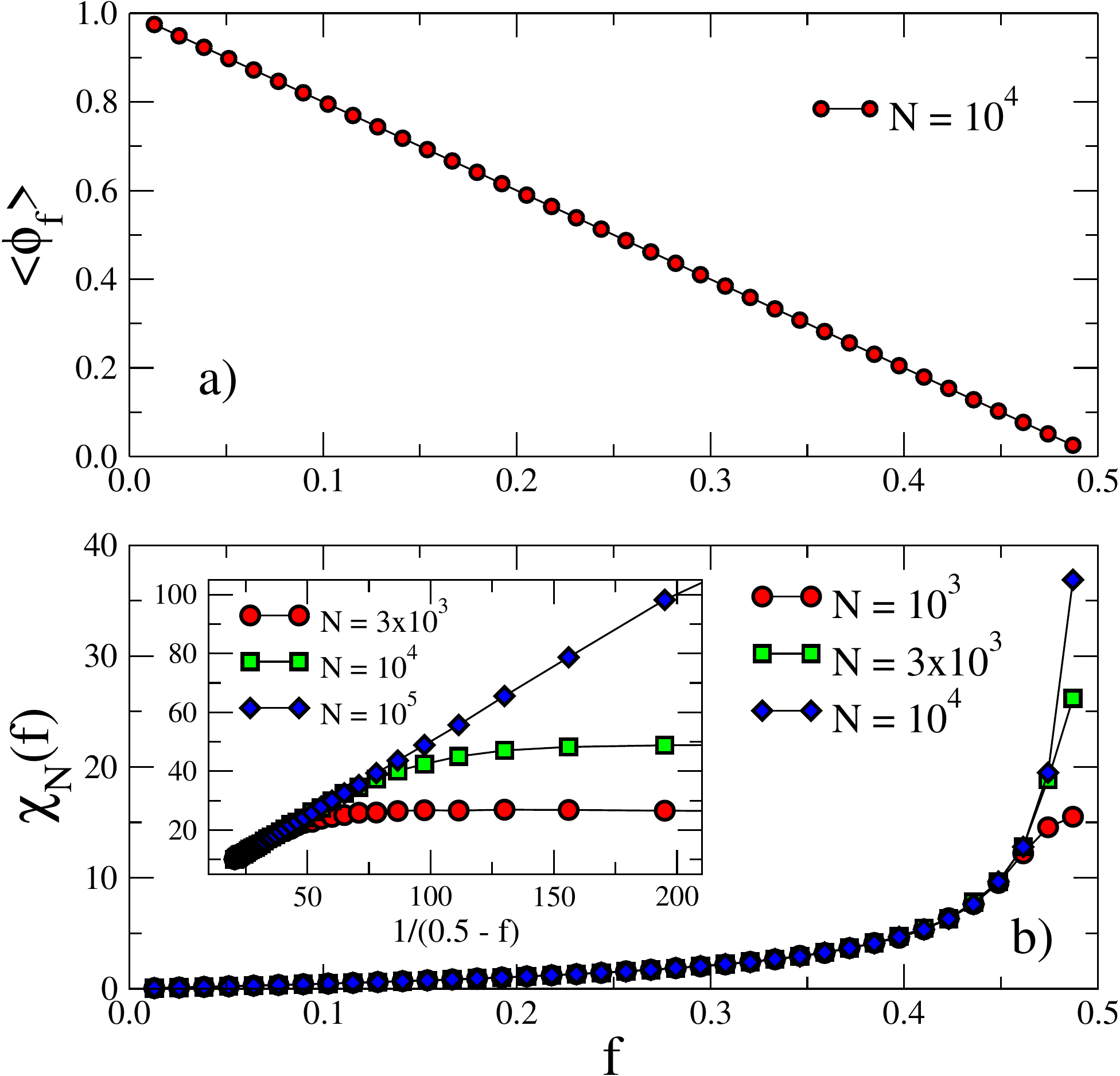}
  \caption{Average order parameter (a) and dynamic susceptibility (b) of
    the majority vote model on fully connected networks of different
    size. Inset of (b): Scaling of the dynamical susceptibility as a
    function of $(0.5 - f)^{-1}$.}
  \label{fig:angularhistogram}
\end{figure}

In Fig.~\ref{fig:angularhistogram} we plot the results of numerical
simulations, averaging over $50000$ Monte Carlo time steps, after
letting the system relax for $10000$ steps.
In Fig.~\ref{fig:angularhistogram}a) we plot the average order parameter
as a function of $f$. The clear linear behavior indicates a value
$f_c = 1/2$ with critical exponent $\beta = 1$. In
Fig.~\ref{fig:angularhistogram}b) we depict instead the dynamics
susceptibility $\chi_N(f)$ as a function of the probability $f$. As we
can see, it tends to diverge close to $f_c = 1/2$, with rounding effects
for small system sizes. In the inset in
Fig.~\ref{fig:angularhistogram}b) we plot $\chi_N(f)$ as a function of
$1/(\frac{1}{2} - f)$. The linear behavior for small values of this
quantity leads to $\chi_N(f) \sim \left( \frac{1}{2} - f\right)^{-1}$,
which combined with  $\chi_N(f) \sim (f_c - f)^{-\gamma -\beta}$ yields
the critical exponent $\gamma = 0$, in agreement with the observations
on the Vicsek model on fully connected graphs.

\begin{figure}[t]
  \includegraphics{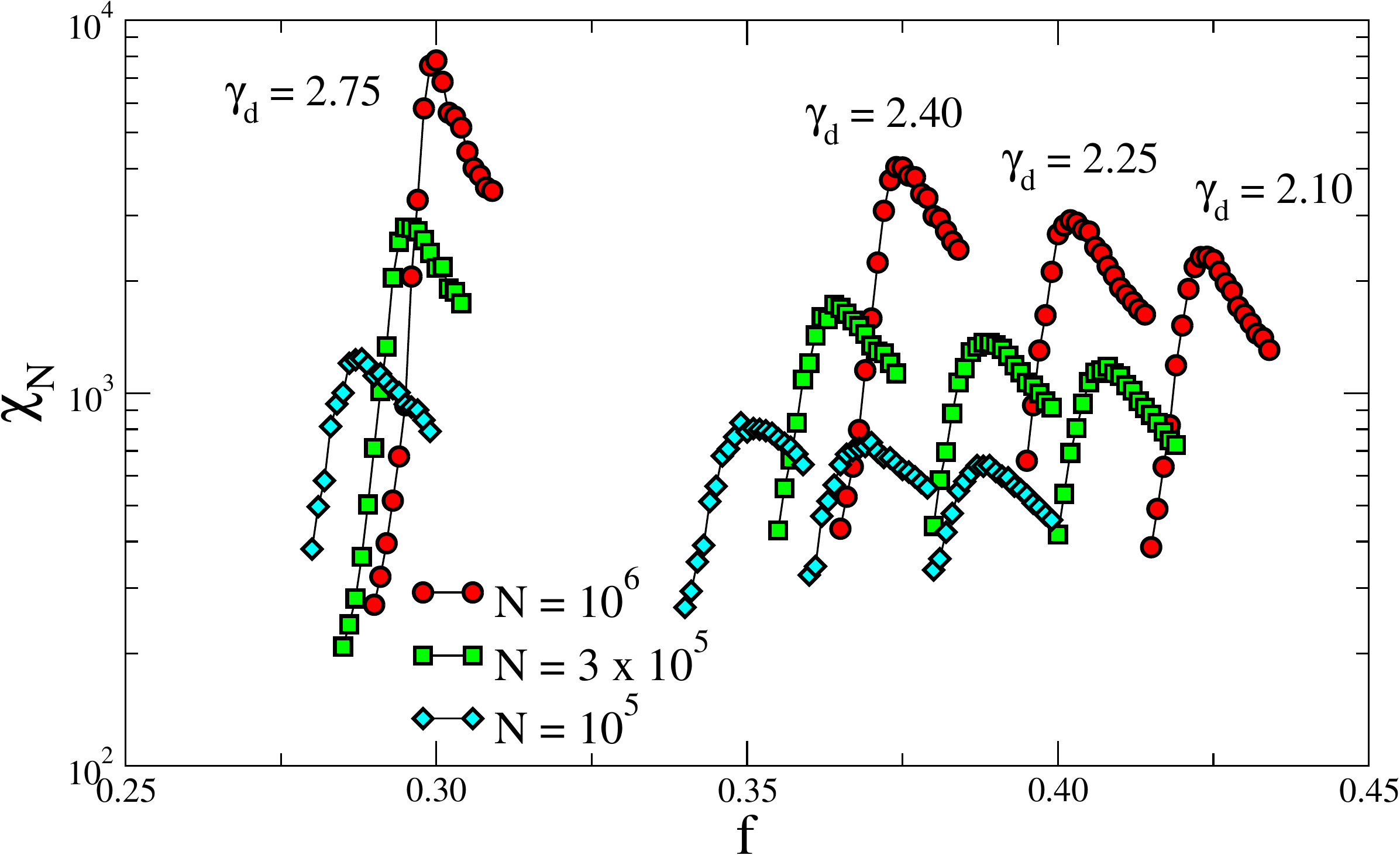}
  \caption{Dynamical susceptibility as a function of $f$ in the majority
    vote model in UCM networks of different size. The groups of
    functions for different $N$ correspond to (from left to right):
    $\gamma_d = 2.75$, $2.40$, $2.25$, and $2.10$.}
  \label{fig:majoritychi}
\end{figure}

\section{Majority vote model on scale-free networks}
\label{sec:majority-vote-model-1}

We consider the majority vote model on scale-free networks of degree
distribution $P(k) \sim k^{-\gamma_d}$, generated using the UCM network
model. In this case, each node takes the majority state of its nearest
neighbors with probability $1-f$, and the opposite state with
probability $f$. In Fig.~\ref{fig:majoritychi} we plot the dynamic
susceptibility $\chi_N(f)$, Eq.~(\ref{eq:4}) as a function of the
probability $f$ in networks of different size $N$ and degree exponent
$\gamma_d$. This function is evaluated averaging over $50000$ Monte
Carlo time steps, after letting the system relax to its steady state. As
we can see, the effective critical point for a given network size,
$f_c(N)$, defined as the position of the peak of the dynamic
susceptibility, appears to tend to a constant for large $\gamma_d$,
while it approaches the limit of large noise $f = 0.5$ in the case of
small $\gamma_d < 5/2$ and large size.  Indeed, the value of the
critical point $f_c$ in the thermodynamic limit, extrapolated from the
finite-size scaling ansatz
\begin{equation}
  \label{eq:5}
  f_c(N) \sim f_c + b N^{-1/\nu},
\end{equation}
leads to a result in agreement with the theoretical prediction in
Ref.~\cite{Chen2015}, namely a value approaching $f_c \simeq 0.5$ for
$\gamma_d < 5/2$, and a value $f_c < 0.5$ for $\gamma_d > 5/2$. 

\begin{figure}[t]
  \includegraphics{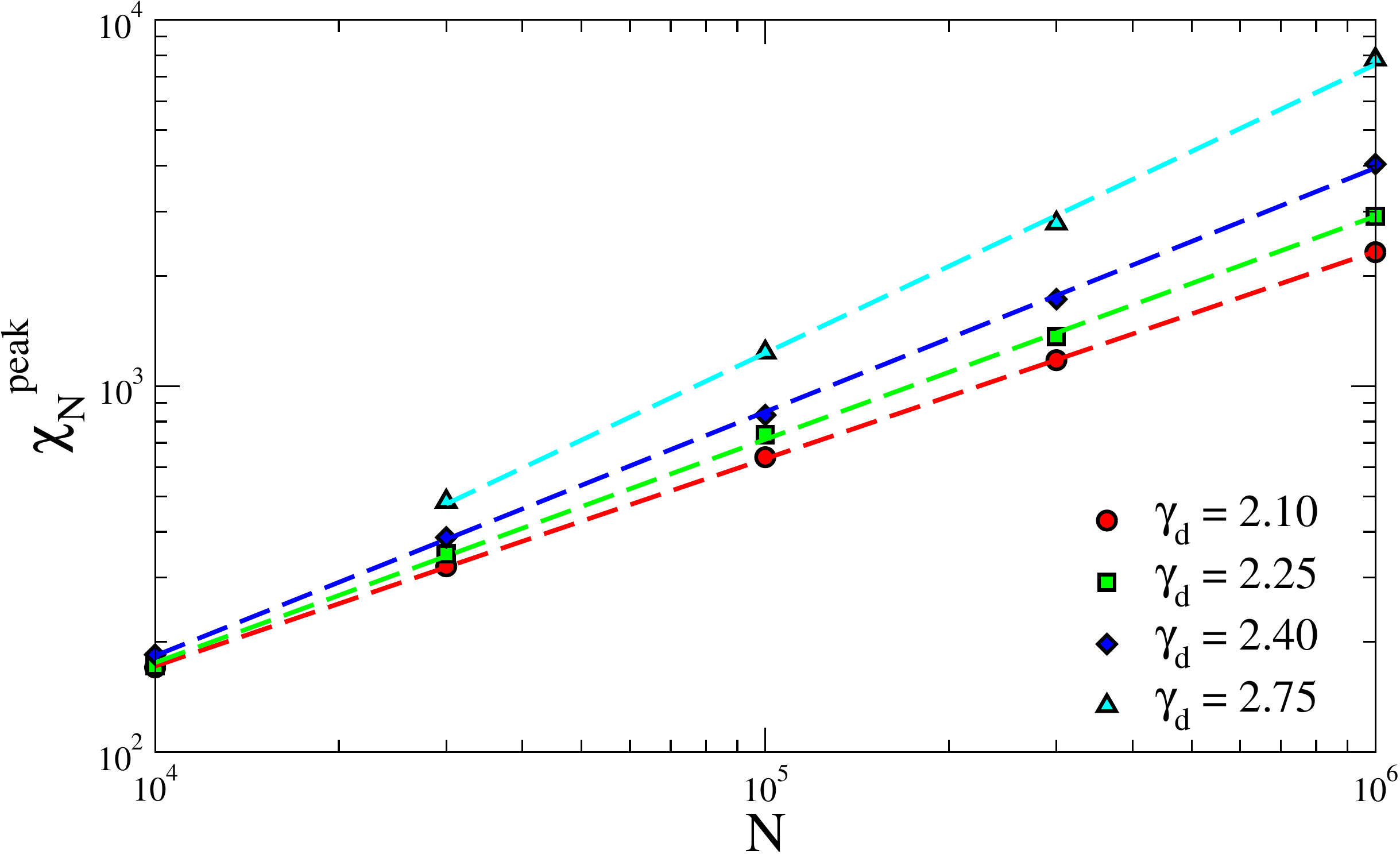}
  \caption{Scaling of the peak of susceptibility with network size for
    different values of $\gamma_d$.}
  \label{fig:majoritydelta}
\end{figure}

In Fig.~\ref{fig:majoritydelta} we plot the height of the peak of the
dynamic susceptibility, which should scale with network size as
\begin{equation}
  \label{eq:6}
  \chi_N^\mathrm{peak} \sim N^{\delta},
\end{equation}
with the exponent $\delta = (\beta + \gamma)/\nu$. By means of a linear
regression in double logarithmic scale, we estimate the values of
$\delta$ for the different degree exponents considered:
$\gamma_d = 2.10$, $\delta = 0.57(1)$; $\gamma_d = 2.25$,
$\delta = 0.61(1)$; $\gamma_d = 2.40$, $\delta = 0.67(2)$;
$\gamma_d = 2.75$, $\delta = 0.78(2)$. These exponents are in excellent
agreement with the ones obtained for the Vicsek model in scale-free
networks (see Table~I in the main paper), and confirm the equivalence of
behavior of the Vicsek model, with a continuous symmetry, and the
majority vote model, with a discrete symmetry, on complex networks with
a heterogeneous degree distribution.

\end{document}